\documentstyle[prl,aps,floats,epsf,color]{revtex}
\begin{document}
\baselineskip=12pt
\def\be{\begin{equation}}
\def\ee{\end{equation}}
\def\bea{\begin{eqnarray}}
\def\eea{\end{eqnarray}}
\def\E{{\rm e}}
\def\bearst{\begin{eqnarray*}}
\def\eearst{\end{eqnarray*}}
\def\peleven{\parbox{11cm}}
\def\peffec{\peight{\bearst\eearst}\hfill\peleven}
\def\pspace{\peight{\bearst\eearst}\hfill}
\def\ptwelve{\parbox{12cm}}
\def\peight{\parbox{8mm}}
\twocolumn
[\hsize\textwidth\columnwidth\hsize\csname@twocolumnfalse\endcsname

\title
{Stochastic $\bf \varphi^4-$Theory in the Strong Coupling Limit}

\author
{N. Abedpour,$^1$ M. D. Niry,$^1$ A. Bahraminasab,$^{2,3}$ A. A.
Masoudi,$^4$\\ J. Davoudi,$^3$ Muhammad Sahimi,$^5$ and  M. Reza
Rahimi Tabar$^{1,6}$}

\vskip 1cm

\address
{$^1$Department of Physics, Sharif University of
Technology, P.O. Box 11365-9161, Tehran, Iran \\
$^2$Department of Physics, Lancaster University, Lancaster, LA1
4YB, UK\\
$^3$International Center for Theoretical Physics, Strada Costiera
11, I-34100 Trieste, Italy\\
$^4$Department of Physics, Alzahra University, Tehran, 19834, Iran\\
$^5$Mork Family Department of Chemical Engineering \& Materials
Science, University of Southern California,\\ Los Angeles,
California 90089-1211\\
$^6$CNRS UMR 6529, Observatoire de la C$\hat o$te d'Azur, BP 4229,
06304 Nice Cedex 4, France  } \maketitle


\begin{abstract}
The stochastic $\varphi^4$-theory in $d-$dimensions dynamically
develops domain wall structures within which the order parameter
is not continuous. We develop a statistical theory for the
$\varphi^4$-theory driven with a random forcing which is white in
time and Gaussian-correlated in space. A master equation is
derived for the probability density function (PDF) of the order
parameter, when the forcing correlation length is much smaller
than the system size, but much larger than the typical width of
the domain walls. Moreover, exact expressions for the one-point
PDF and all the moments $\langle\varphi^n\rangle$ are given. We
then investigate the intermittency issue in the strong coupling
limit, and derive the tail of the PDF of the increments
$\varphi(x_2) - \varphi(x_1)$. The scaling laws for the structure
functions of the increments are obtained through numerical
simulations. It is shown that the moments of field increments
defined by, $C_b=\langle |\varphi(x_2)-\varphi(x_1)|^b\rangle$,
behave as $|x_1-x_2|^{\xi_b}$, where $\xi_b=b$ for $b\leq 1$, and
$\xi_b=1$ for
$b\geq1$.\\

\end{abstract}
\hspace{.3in}
\newpage
]

{
\section{Introduction}

Thirty five years ago Wilson and Fisher \cite{01} emphasized the
relevance of the $\varphi^4$-theory to understanding the critical
phenomena. Since then, the theory has become one of the most
appealing theoretical tools for studying the critical phenomena in
a wide variety of systems in statistical physics. In the strong
coupling limit, the $\varphi^4$-theory develops domain walls, a
phenomenon which is of great interest in the classical and quantum
field theories [2-10]. The dynamical $\varphi^4-$theory - what is
usually referred to as the time-dependent Ginzburg-Landau (GL)
theory - provides a phenomenological approach to, and plays an
important role in, understanding dynamical phase transitions and
calculating the associated dynamical exponent [11-16]. The
time-dependent GL theory for superconductors was presented
phenomenologically only in 1968 by Schmid \cite{Schmid} (and
derived from microscopic theory shortly thereafter \cite{Gorkov}),
when the first modulational theory was derived in the context of
Rayleigh-Benard convection \cite{Newell,Segel}. Moreover, the GL
equation with an additional noise term has been studied
intensively as a model of phase transitions in equilibrium
systems; see, for example, \cite{Hohenberg}.

In the present paper we consider the stochastic $\varphi^4$-theory
in the strong coupling limit. This limit is singular in the sense
that, the equation that describes the dynamics of the system
develops singularities. Therefore, starting with a smooth initial
condition, the domain-wall singularities are dynamically developed
after a finite time. At the singular points the field
$\varphi(x,t)$ is not continuous. We derive master equations for
the joint probability density functions (PDF) of $\varphi$ and its
increments in $d$ dimensions. It is shown that in the stationary
state, where the singularities are fully developed, the relaxation
term in the strong coupling limit leads to an unclosed term in the
PDF equations.

Using the boundary layer method \cite{BO,16}, we show that the
unclosed term makes no finite contribution (anomaly) in the strong
coupling limit, and derive the PDF of $\varphi$ and its moments,
$\langle\varphi^n\rangle$, in the same limit. We also investigate
the scaling behavior of the moments of the field's increments
defined by, $\delta\varphi=\varphi(x_2)-\varphi(x_1)$, and show
that when $|x_2-x_1|$ is small, fluctuations of the $\varphi$
field have a bi-fractal structure and are intermittent. The
intermittency implies that the structure function defined by,
$C_b=\langle|\varphi(x_2)-\varphi(x_1)|^b\rangle$, scales as
$|x_2-x_1|^{\zeta_b}$, where $\zeta_b$ is a nonlinear function of
$b$. It is also shown numerically that the moments of the field's
increments, $\langle|\varphi(x_2)-\varphi(x_1)|^b\rangle$, behave
as, $|x_2-x_1|^{\xi_b}$, where $x_b=b$ for $b\leq 1$, and
$\xi_b=1$ for $b\geq1$.

The rest of this paper is organized as follows. In the next
section we present the model that we wish to study, and describe
some of its properties by solving it numerically. In III and IV we
derive master equations for the order parameter of the model, and
for the field's increments and its PDF tail. The numerical
simulations for extracting the scaling exponents are described in
V. The paper is summarized in VI, while the Appendices  provide
some technical details of the work that we present in the main
part of the paper.

\section{The Model and the Coupling Constant}

The standard GL $\varphi^4$-theory describes a second-order phase
transition in any system with a one-component order parameter
$\varphi(x)$ and the $\varphi\to -\varphi$ symmetry in a zero
external field. The theory is described by the following action,
\begin{equation}
S=\int\left[-\frac{k}{2}(\nabla\varphi)^2+\frac{\tau}{2!}\varphi^2
-\frac{g}{4!}\varphi^4\right]d^dx\;,
\end{equation}
where $\tau=T-T_c$, with $T_c$ being the critical temperature, and
$k$ is the diffusion coefficient. We consider the case in which
$\tau > 0$. For $d\geq 2$, the critical temperature $T_c$ is
finite, while in one dimension (1D), $T_c=0$. The parameter $g$
characterizes the strength of the fluctuation interaction, or the
coupling constant. The equation of motion is given by,
\begin{equation}
k\nabla^2\varphi+\tau\varphi-\frac{g}{6}\varphi^3=0.
\end{equation}

The critical dynamics of the system is described by a stochastic
equation of a particular form - the Langevin equation (for a
comprehensive review on Langevin type equations see \cite{Risken}
)- given by
\begin{equation}
\frac{\partial\varphi}{\partial
t}=k\nabla^2\varphi+\tau\varphi-\frac{g}{6} \varphi^3+\eta({\bf
x},t)\;,
\end{equation}
where $\eta({\bf x},t)$ is a Gaussian-distributed noise with zero
average and the correlation function,
\begin{equation}
\langle\eta({\bf x},t)\eta({\bf x}',t')\rangle=D_0D({\bf x}-{\bf
x}') \delta(t-t')\;,
\end{equation}
with $D({\bf x}-{\bf x}')$ being an arbitrary smooth function.
Typically, the spatial correlation of the forcing term is
considered to be a delta function in order to mimic short-range
correlations. Here, though, the spatial correlation is defined by
\begin{equation}\label{}
D({\bf x}-{\bf
x}')=\frac{1}{({\pi\sigma^2})^{d/2}}\exp\left[-\frac{({\bf x}-
{\bf x}')^2}{\sigma^2}\right]\;,
\end{equation}
where $\sigma\ll L$ endows a short-range character to the random
forcing. It is useful to rescale Eq. (3) by writing,
$\varphi'=\varphi/\varphi_0$, $x'=x/x_0$, and $t'=t/t_0$. If we
let $t_0=1/\tau$, $\varphi_0=(6\tau/g)^{1/2}$, and, $x_0=
[(d_0g)/(6\tau^2)]^{1/d}$, all the parameters are eliminated in
Eq. (3) except for, $k'=[6/(D_0g)]^{2/d}\tau^{4/d-1}k$, and one
finds that
\begin{equation}\label{eqphi4}
\frac{\partial\varphi}{\partial
t}=k'\nabla^2\varphi+\varphi-\varphi^3 +\eta({\bf x},t)\;,
\end{equation}
where $k'$ is now the effective coupling constant of the theory.
The weak and strong coupling limits of the theory are then
defined, respectively, by, $k'\to \infty$ and $k'\to 0$. In the
weak coupling limit one can use numerical simulations and the
Feynman diagrams to calculate the critical exponents. On the other
hand, to solve the problem in the strong coupling limit we need
other techniques to derive the stochastic properties of the
fluctuation field [16]. The nonlinearity of Eq. (6) in the strong
coupling limit gives rise to the possibility of formation of
singularity in a finite time. This means that there is a
competition between the smoothing effect of diffusion (the
Laplacian term) and the $\varphi^3$ term. Let us now describe the
main properties of the GL theory in the limit, $k'\to 0$.

i) The unforced GL model [$\eta({\bf x},t)=0$], with given initial
conditions, develops singularities in any spatial dimension. In
one spatial dimension (1D) the singularities are developed in a
finite time $t_c$ as $k'\to 0$. At such singular points the field
$\varphi$, representing an order parameter, is not continuous. In
2D the unforced GL model develops domain walls, characterized by
singular lines with finite lengths (that depend on the initial
condition). Under these conditions, the field $\varphi$ is
discontinuous when crossing the singular lines. In three and
higher dimensions the structure of the singularities can be more
complex. For example, in 3D the singularities are domain walls
where the field $\varphi$ is discontinuous.

In Figs. 1 we show the time evolution of the order parameter
$\varphi$ of the unforced GL model in 2D, in the limit $k'\to 0$
[Eq. (6) with $\eta=0$]. We have used the finite-element method to
numerically solve the Langevin equation with $k'\to 0$ and the
initial condition, $\varphi(x,y,0)=\sin x\sin y$. Such initial
conditions are typical, and were used only for simplicity. The
time scale for reaching the singularity is of the order of
$k'^{-1/2}$. As Figs. 1a and 1b indicate, it is evident that at
times $t<t_c$ (in the limit, $k'\to 0$) the $\varphi$ field is
continuous. At $t=t_c$ the $\varphi$ field becomes singular; see
Fig. 1c.

(ii) Similarly, for a forcing term which is white noise in time
and smooth in space, singularities are developed in any spatial
dimension in the strong coupling limit and in a finite time,
$t_\eta =t_{c,\eta}$, as $k'\to 0$. For example, in 2D the
boundaries of the domain walls are smooth curves. In Fig. 2 we
demonstrate the time evolution of the order parameter $\varphi$ of
the forced GL model in 2D in the limit $k'\to 0$. Starting from a
smooth initial condition, as shown in Figs. 2a and 2b, it is
evident that for times $t<t_{c,\eta}$ the $\varphi$ field is
continuous. At $t=t_{c,\eta}$ the field becomes singular; see Fig.
2c.

\begin{figure}[t]\label{fig01}
\epsfxsize=6truecm\epsfbox{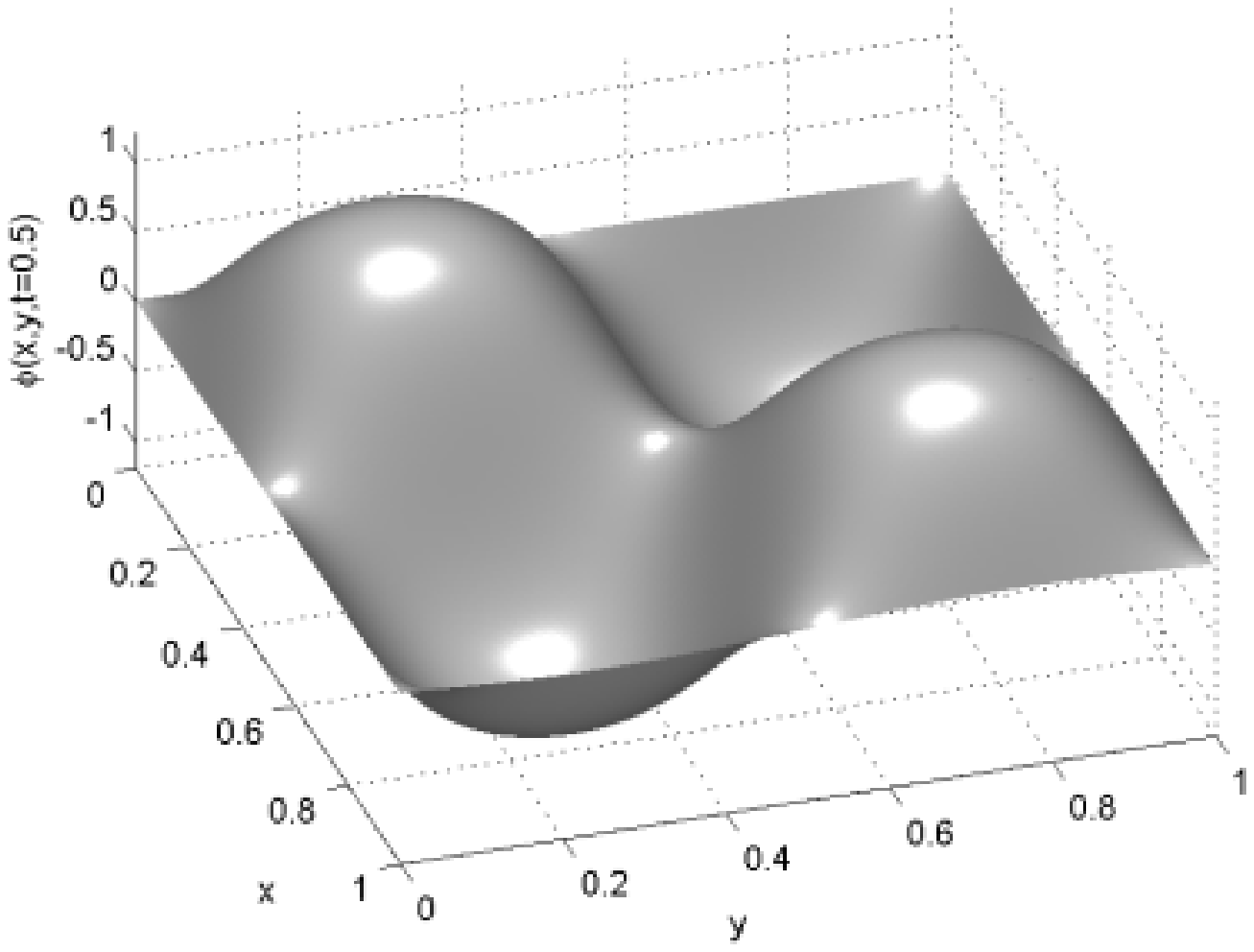}
\epsfxsize=6truecm\epsfbox{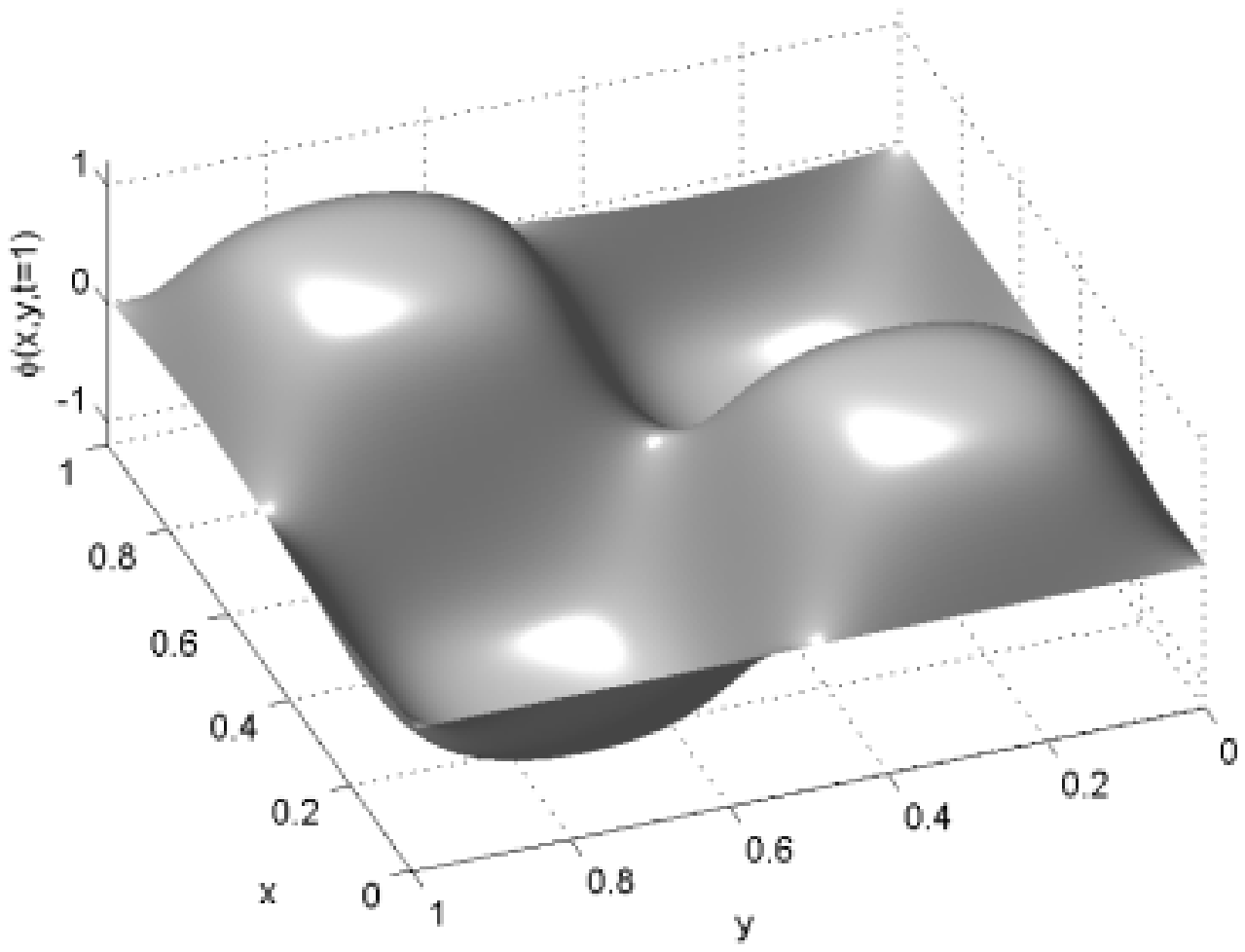}
\epsfxsize=6truecm\epsfbox{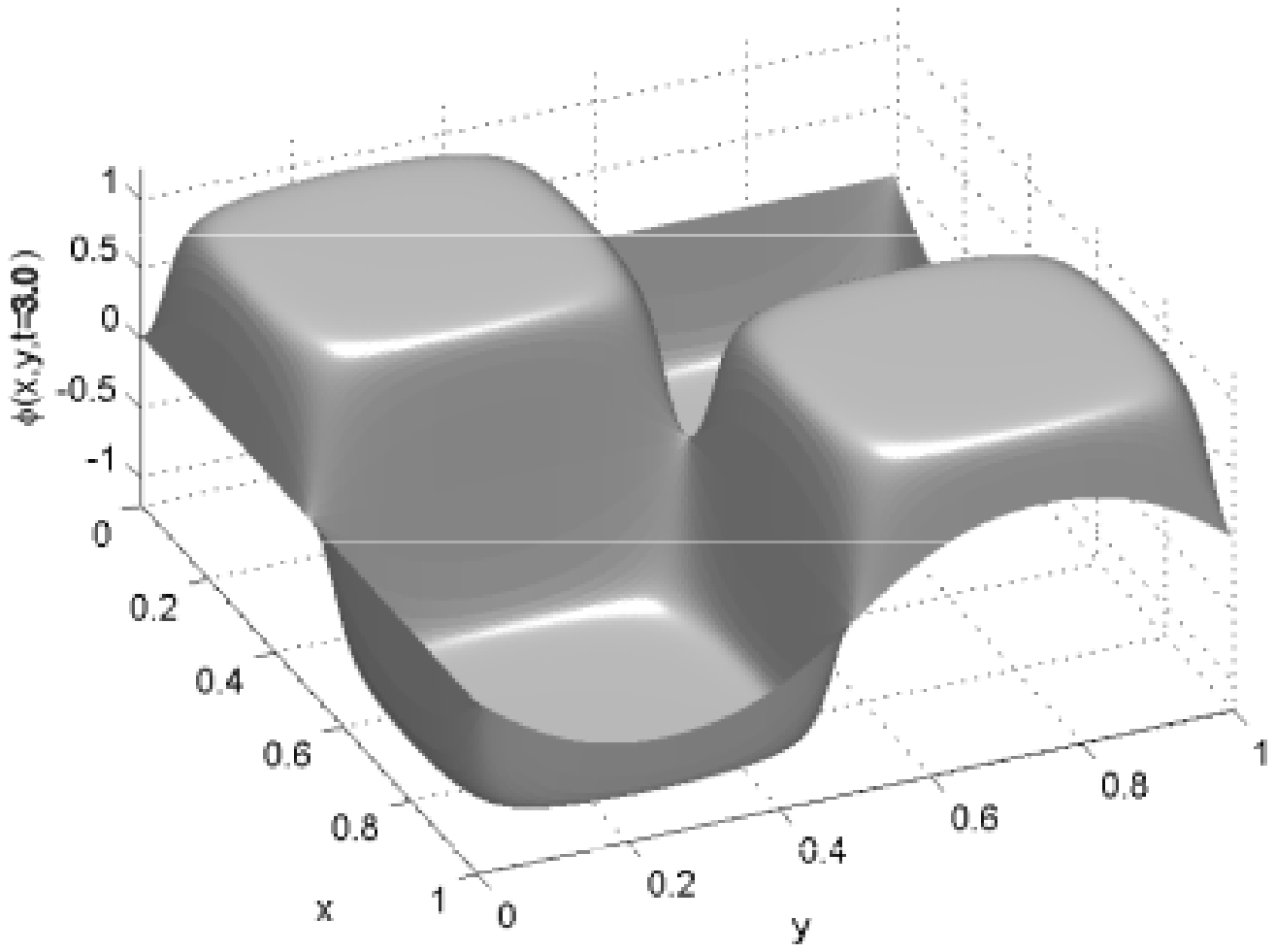} \caption{Snapshots of the order
parameter $\varphi$ at various times in the 2D unforced
$\varphi^4$-theory in the strong coupling limit, $k'\to 0$. Top
two figures show the $\varphi$ field before the singularity
develops. In the bottom figure, which is for time scales greater
than the time at which singularity develops, the $\varphi(x,y)$
field is not continuous. The initial condition is,
$\varphi(x,y,0)=\sin x\sin y$.}
\end{figure}}

\begin{figure}[t]\label{fig02}
\epsfxsize=6truecm\epsfbox{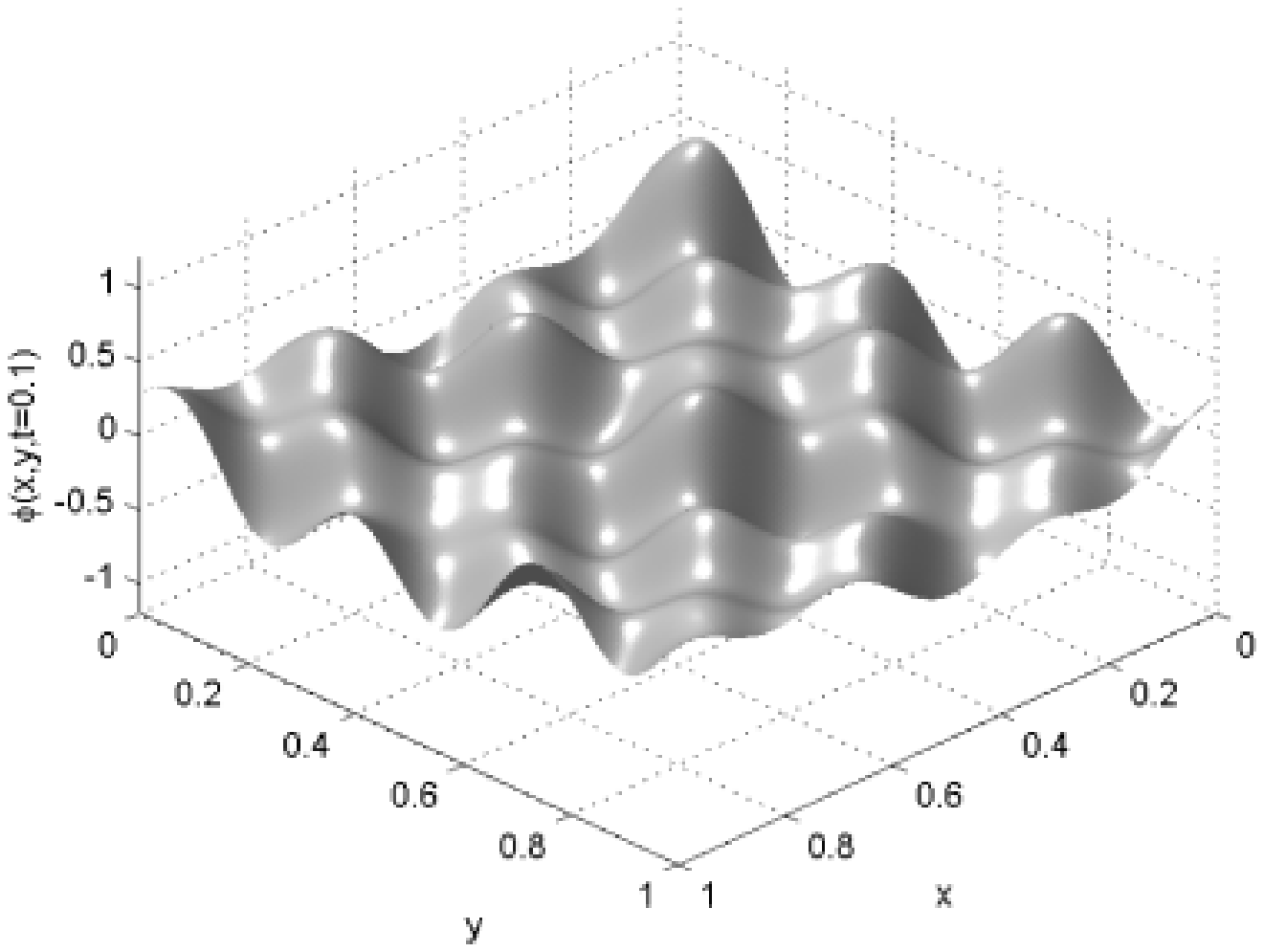}
\epsfxsize=6truecm\epsfbox{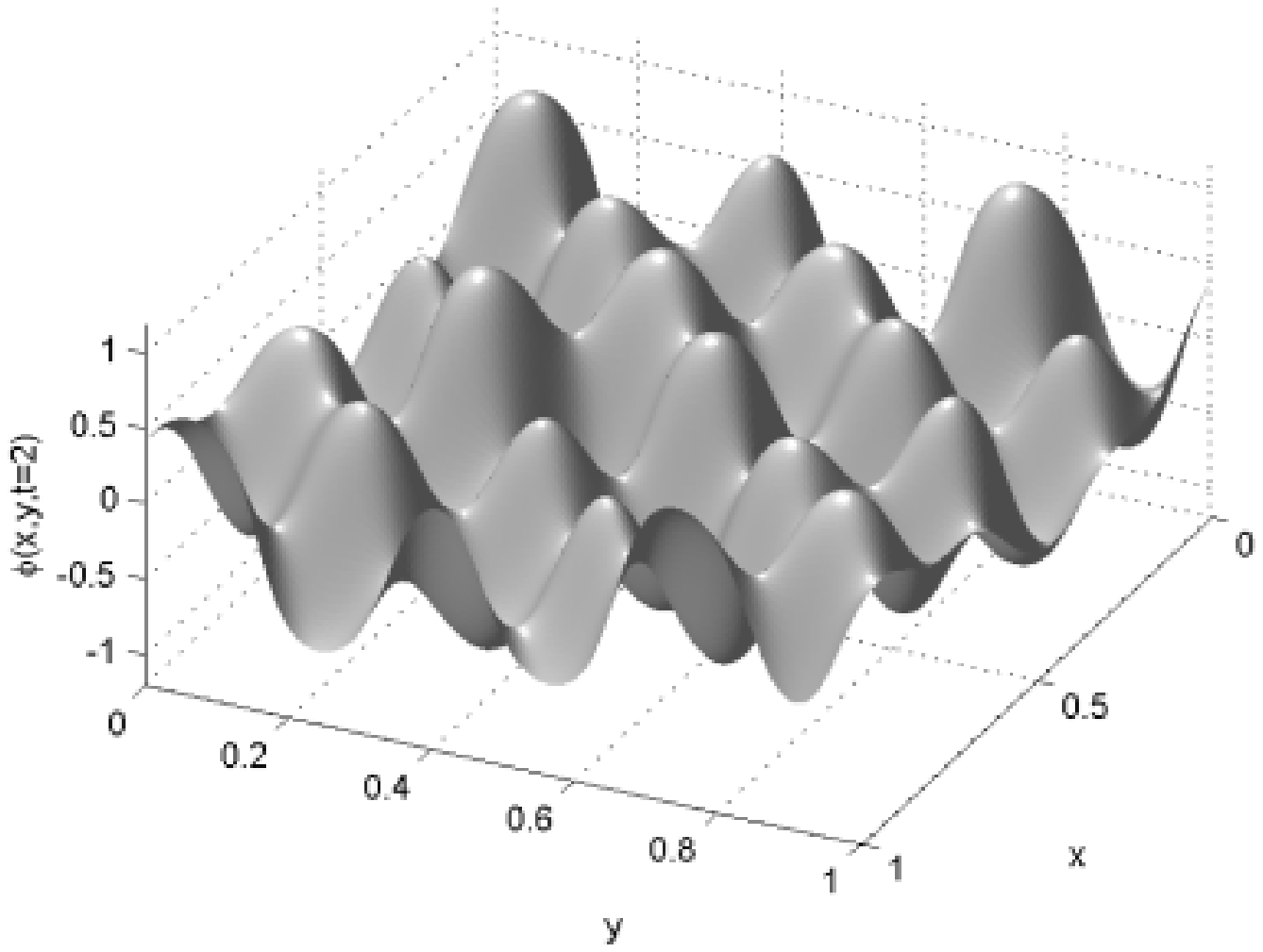}
\epsfxsize=6truecm\epsfbox{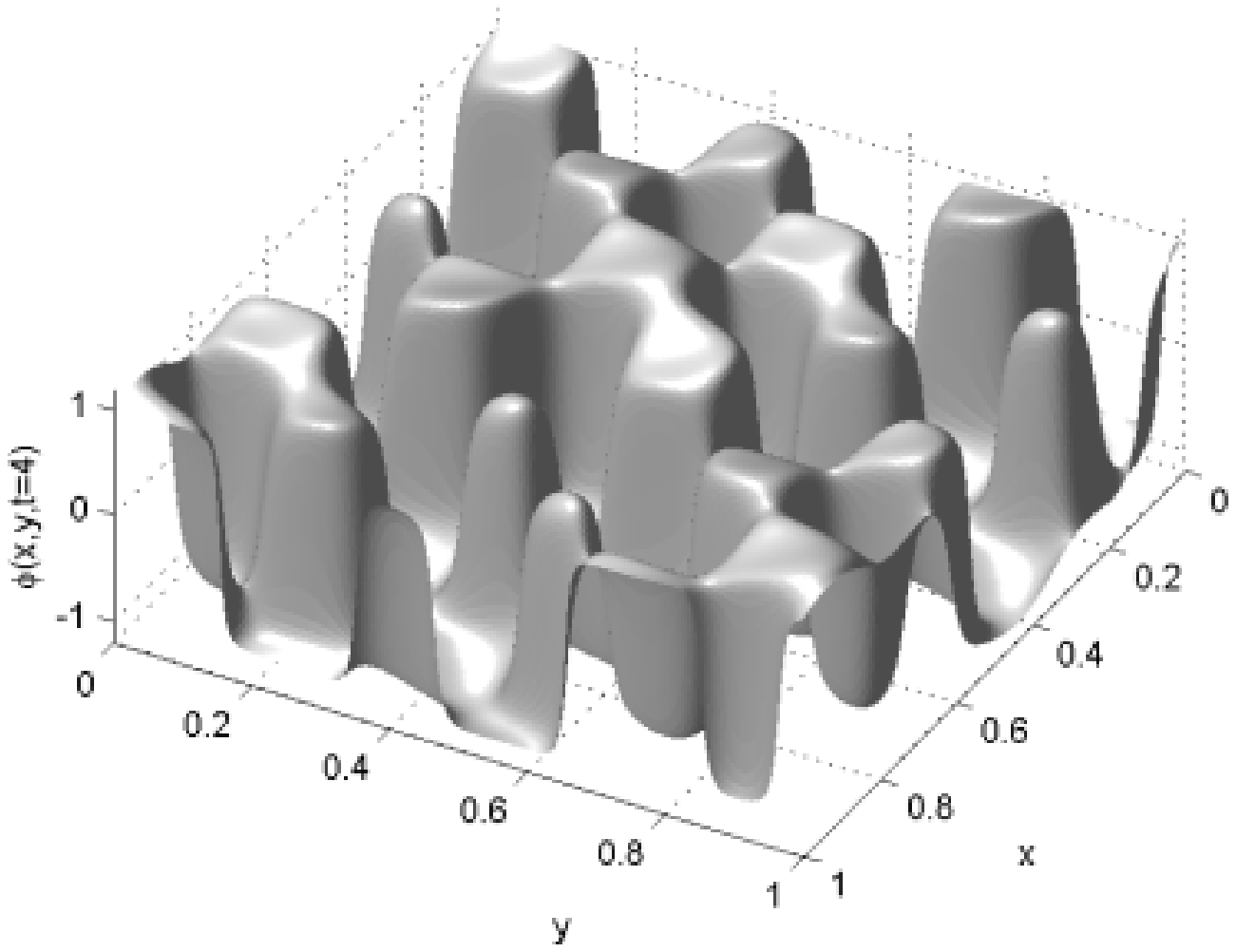} \caption{Snapshots of the order
parameter $\varphi$ of a randomly-driven 2D $\varphi^4$-theory in
the strong coupling limit. In the simulations the relation between
the forcing length scale $\sigma$ and the sample size $L$ is,
$\sigma \simeq L/3$. The forcing strength $D_0$ is 0.1.}
\end{figure}

\section {Master Equation of the Order Parameter}

In this section we derive a master equation to describe the time
evolution of the PDF $P(\varphi,t)$ of the order parameter
$\varphi$. Defining a one-point generating function by,
$Z(\lambda)=\langle\Theta\rangle$, where $\Theta$ is defined by,
$\Theta=\exp[-i\lambda\varphi(x,t)]$. Using Eq. (6), the time
evolution of $Z$ is governed by
\begin{eqnarray}
Z_t=-i\lambda
k'\langle\nabla^2\varphi\exp[-i\lambda\varphi(x,t)]\rangle
-i\lambda\langle\varphi\Theta\rangle\nonumber\\+i\lambda\langle\varphi^3\Theta
\rangle-i\lambda\langle\eta\Theta\rangle-\lambda^2k(0)Z\;,
\end{eqnarray}
where, $k(x)=D_0D({\bf x})$, and we have invoked Novikov's theorem
(see Appendix I), which is expressed via the relation,
\begin{equation}\label{nov}
\langle\eta\exp[-i\lambda\varphi(x,t)]\rangle=-i\lambda k(0) Z\;.
\end{equation}
Now, using the identities,
$-i\lambda\langle\varphi\exp[-i\lambda\varphi(x,t)]
\rangle=\lambda Z_\lambda$, and
$-i\lambda\langle\varphi^3\exp[-i\lambda \varphi({\bf
x},t)]\rangle=\lambda Z_\lambda\lambda\lambda$, the generating
function $Z$ satisfies the following unclosed master equation
\begin{eqnarray}\label{ZFG}
Z_t=-i\lambda
k'\langle\nabla^2\varphi\exp[-i\lambda\varphi(x,t)]\rangle
+\lambda Z_\lambda+\lambda Z_{\lambda\lambda\lambda}\nonumber\\
-\lambda^2 k(0)Z\;.
\end{eqnarray}
The $-i\lambda
k'\langle\nabla^2\varphi\exp[-i\lambda\varphi(x,t)]\rangle$ term
of Eq. (\ref{ZFG}) is the only one which is not closed with
respect to $Z$. The PDF of order parameter $P(\varphi)$ is
constructed by Fourier transforming the generating function $Z$:
\begin{equation}
P(\varphi,t)=\int\frac{d\lambda}{2\pi}\exp(i\lambda\varphi)Z(\lambda,t)\;.
\end{equation}
Thus,
\begin{eqnarray}
P_{t}&=&-[(\varphi-\varphi^3)P]_{\varphi}+k(0)P_{\varphi\varphi}\nonumber\\
&-&ik'\int\frac{d\lambda}{2\pi}\lambda\exp(i\lambda\varphi)
\langle\nabla^2\varphi\exp[-i\lambda\varphi(x,t)]\rangle\;.
\end{eqnarray}
It is evident that the governing equation for $P(\varphi,t)$ is
also not closed.

Let us now use the boundary layer technique to prove that the
unclosed term [the last term of Eq. (11)] makes, in the strong
coupling limit, no contribution to the governing equation for the
PDF\cite{BO,16}. We consider two different time scales in the
limit, $k'\to 0$. (i) Early stages before developing the
singularities ($t<t_{c,\eta}$), and (ii) in the regime of
established stationary state with fully-developed sharp
singularities ($t \geq t_{c,\eta}$).

In regime (i), ignoring the relaxation term in the governing
equation for the PDF, one finds, in the limit $k'\to 0$, the exact
equation for the time evolution of the PDF for the order parameter
(see below for more details). In contrast, the limit $k'\to 0$ is
singular in regime (ii), leading to an unclosed term (the
relaxation term) in the equation for the PDF. However, we show
that the unclosed term scales as $k'^{1/2}$, implying that this
term, in the strong coupling limit, makes no finite contribution
or anomaly to the solution of Eq. (11). It is known for such time
scales (the stationary state) that the $\varphi$-field, which
satisfies the Langevin equation, gives rise to discontinuous
solutions in the limit, $k'\to 0$. Consequently, for finite
$\sigma$ the singular solutions form a set of points where the
domain walls are located, and are continuously connected. We
should note that $k' \varphi_{xx}$, in the limit $k'\to 0$, is
zero at those points at which there is no singularity. Therefore,
in the limit $k'\to 0$ only small intervals around the walls
contribute to the integral in Eq. (11). Within these intervals, a
boundary layer analysis may be used for obtaining accurate
approximation of $\varphi({\bf x},t)$.

Generally speaking, the boundary layer analysis deals with
problems in which the perturbations are operative over very narrow
regions, across which the dependent variables undergo very rapid
changes. The narrow regions, usually referred to as the domain
walls, frequently adjoin the boundaries of the domain of interest,
due to the fact that a small parameter ($k'$ in the present
problem) multiplies the highest derivative. A powerful method for
treating the boundary layer problems is the method of matched
asymptotic expansions. The basic idea underlying this method is
that, an approximate solution to a given problem is sought, not as
a single expansion in terms of a single scale, but as two or more
separate expansions in terms of two or more scales, each of which
is valid in some part of the domain. The scales are selected such
that the expansion as a whole covers the entire domain of
interest, and the domains  of validity of neighboring expansions
overlap. In order to handle the rapid variations in the domain
walls' layers, we define a suitable magnified or stretched scale
and expand the functions in terms of it in the domain walls'
regions.

For this purpose, we split $\varphi$ into a sum of inner solution
near the domain walls and an outer solution away from the
singularity lines, and use systematic matched asymptotics to
construct a uniform approximation of $\varphi$. For the outer
solution, we look for an approximation in the form of a series in
$k'$,
\begin{equation}
\varphi=\varphi^{\rm out}=\varphi_0+k'\varphi_1+O(k'^2)\;,
\end{equation}
where $\varphi_0$ satisfies the following equation
\begin{equation}
\varphi_{0t}=\varphi_0-\varphi_0^3+\eta({\bf x},t)\;.
\end{equation}
Indeed, $\varphi_0$ satisfies Eq. (6) with $k'=0$. Far from the
singular points or lines, the PDF of $\varphi_0$ satisfies the
Fokker-Planck equation, with the drift and diffusion coefficients
being, $D^{(1)}(\varphi_0,t)=\varphi_0- \varphi_0^3 $, and
$D^{(2)}=k(0)$, respectively. Reference [16] gives the solution of
the time-dependent Fokker-Planck equation with such drift and
diffusion coefficients. At long times and in the area far from the
singular points or lines, the PDF of $\varphi_0$ will have two
maxima at $\pm 1$. This means that we are dealing with the smooth
areas in Fig. 2c in the stationary state.

In order to deal with the inner solution around the domain walls,
we consider the $x$ component {\it normal} to the domain wall or
singularity line, and decompose the operator $\nabla^2$ as
$\partial_{xx}+\nabla_{d-1}^2$. In the strong coupling limit,
$k'\to 0$, the term $\nabla_{d-1}^2\varphi$ makes no contribution
to the PDF equation, whereas the term $\partial_{xx}\varphi$ is
singular. To derive the long-time solution of Eq. (6), we rescale
$x$ to $z\equiv\frac{x}{\sqrt{2k'}}$ and suppose that complete
solution of Eq. (6) has the form, $\varphi(z,t)=f(z,t)+\tanh(z)$.
All the effects of the initial condition and time-dependence of
$\varphi(x,t)$ will then be contained in $f(z,t)$. We now rewrite
Eq. (6) with the new variables to obtain
\begin{eqnarray}\label{eqnnphi4}
\partial_t f(z,t)=&&\frac{1}{2}\partial_{zz}f(z,t)+f(z,t)-f^3(z,t)\nonumber\\
&&
-3f(z,t)\tanh(z)[f(z,t)+\tanh(z)]\nonumber\\&&+\sqrt[4]{2k'}\eta(z,t)\;.
\end{eqnarray}
The last term of Eq. (14) is zero in the limit $k'\to 0$.
Multiplying Eq. (14) by $f(z,t)$ and integrating over $z$, one
finds that,
\begin{eqnarray}
&& \partial_t\int dz f^2(z,t)=- \int dz (\partial_{z}f(z,t))^2 + 2
\int dz f^2(z,t)\nonumber\\&&- 2 \int dz f^4(z,t) -6\int dz
f^3(z,t)a(z)  \nonumber \\ &&- 6\int dz a^2(z)f(z,t)^2\;,
\end{eqnarray}
where $a(z)=\tanh(z)$. We show in Fig. (3) the time variations of
$\int dz f^2(z,t) $ verses $t$ with different types of initial
conditions. The results show that $\int dz f^2(z,t)$ vanishes at
long times. Therefore, $\varphi(z,t)\to\tanh(z)$, in the limit of
a stationary state. Let us now compute the contribution of the
unclosed term in Eq. (11) in the stationary state, i.e.,
\begin{eqnarray}
&& -k'\int \frac{d\lambda}{2\pi}i\lambda
e^{i\lambda\varphi}\langle\nabla^2\varphi
e^{-i\lambda\varphi(x,t)}\rangle\cr \nonumber\\&& = -k'\left(\int
\frac{d\lambda}{2\pi} e^{i\lambda\varphi}\langle\nabla^2\varphi
e^{-i\lambda\varphi(x,t)}\rangle\right)_\varphi\nonumber\\&&
=-k'\langle\nabla^2\varphi\delta[\varphi-\varphi(x,t)]\rangle_\varphi
\end{eqnarray}
In the second line of Eq. (16), we have replaced $i\lambda$ with
differentiation with respect to $\varphi$ and in the third line
the integration of $\lambda$ has been carried through. Now,
assuming ergodicity, the term
$k'\langle\nabla^2\varphi\delta(\varphi-\varphi(x,t))\rangle$ is
converted to,
\begin{equation}
=-k'\lim_{V\to\infty}\frac{1}{V}\int_V dxdv_{d-1}\nabla^2\varphi
\delta[\varphi-\varphi(x,t)]\;.
\end{equation}
In the limit $k'\to 0$, only at points where we have singularity
the above term is not zero. Therefore, we approach the domain
walls' regions as
\begin{eqnarray}
= -k'\lim_{V\to\infty}\frac{1}{V}\sum_j\int_{\Omega_j}
dxdv_{d-1}(\varphi_{xx}+\nabla^{2}_{d-1}\varphi)\nonumber\\
\times\delta[\varphi-\varphi(x,t)]\;,
\end{eqnarray}
where $\Omega_j$ is the space close to the domain walls.
Therefore, Eq. (16) is written as
\begin{equation}
= -k'\lim_{V\to\infty}\frac{1}{V}\sum_j\int_{\Omega_j} dxdv_{d-1}
\varphi_{xx}\delta[\varphi-\varphi(x,t)]\;.
\end{equation}

Changing the variables from $x$ to $z$ and integrating over
$dv_{d-1}$, we have
\begin{eqnarray*}
=
-k'\lim_{V\to\infty}\frac{V_{d-1}}{V}\sum_j\int_{-\infty}^{+\infty}
\epsilon dz \frac{1}{\epsilon^2}\varphi_{zz}\delta[\varphi-\varphi(z,t)]\\
= -\frac{k'}{\epsilon}\lim_{V\to\infty}\frac{V_{d-1}}{V}\sum_j
\int_{-\infty}^{+\infty}dz\varphi_{zz}\delta[\varphi-\varphi(z,t)]\\
\end{eqnarray*}
where $\epsilon=(2k')^{1/2}$. Assuming statistical homogeneity,
one finds,
\begin{equation}
= -\frac{k'}{\epsilon}\lim_{V\to\infty}\frac{N\times V_{d-1}}{V}
\int_{-\infty}^{+\infty}dz \varphi_{zz}\delta[\varphi-\varphi(z,t)]\\
\end{equation}
where $N$ is number of singular lines. The quantity $NV_{d-1}/V$
is the density of the singular lines, and
$k'/\epsilon=(k'/2)^{1/2}$. In the limit, $V\to \infty$, we denote
the density of the singularities by $\rho$. Therefore,
\begin{equation}
=-(k'/2)^{1/2}\rho\int_{-\infty}^{+\infty}dz \varphi_{zz}
\delta[\varphi-\varphi(z,t)]\;.
\end{equation}
Now, by changing the integration variable form $z$ to $\varphi$,
we can determine the integral exactly. We find that,
\begin{equation}
\int_{-\infty}^{+\infty}dz
\varphi_{zz}\delta[\varphi-\varphi(z,t)] =\int_{-1}^{+1} d\varphi
\frac{\varphi_{zz}}{\varphi_{z}}\delta[\varphi-\varphi(z,t)]\;.
\end{equation}
Using Eq. (6) in the limit, $t\to\infty$, we determine
$\varphi_{zz}$ and $\varphi_z$ in terms of $\varphi$. Multiplying
Eq. (6) by $\varphi_z$ and integrating over $z$, we obtain,
\begin{equation}
\varphi_z^2=\frac{1}{2}\varphi^4-\varphi^2+C\;,
\end{equation}
where $C$ is an integration constant. In the limit,
$z\to\pm\infty$, $\varphi_z=\varphi=\pm 1$. Therefore, $C=1/2$,
and $\varphi_{zz}/\varphi_z$ is written as,
\begin{eqnarray*}
\frac{\varphi_{zz}}{\varphi_z}=\frac{\varphi^3-\varphi}{\sqrt{|\frac{1}{2}
\varphi^4-\varphi^2+\frac{1}{2}|}}\\
=\frac{\sqrt{2}\varphi(\varphi^2-1)}{|\varphi^2-1|}\\
=\sqrt{2}\varphi\times sign(\varphi^2-1)\;.
\end{eqnarray*}
The integral in Eq. (22) is now given by,
\begin{eqnarray*}
&&\int_{-1}^{+1}d\varphi\frac{\varphi_{zz}}{\varphi_z}
\delta[\varphi-\varphi(z,t)]\\
&=&\int_{-1}^{+1} d\varphi(z) \sqrt{2}\varphi(z)\times
sign(\varphi(z)^2-1)\delta(\varphi-\varphi(z,t))\\
&=&\sqrt{2}\varphi \hskip 0.1cm sign(\varphi^2 -1)
\theta(1-\varphi^2).
\end{eqnarray*}
Now, the unclosed term in Eq. (16) is written as
\begin{eqnarray*}
&&-k'\langle\nabla^2\varphi\delta[\varphi-\varphi(x,t)]\rangle_{\varphi}=\\
&=& -(k'/2)^{1/2}\rho\{A+2\sqrt{2}\varphi^2\delta(\varphi^2-1)\},
\end{eqnarray*}
where, $A=2\sqrt{2}\theta(1-\varphi^2)\theta(\varphi^2-1)-\sqrt{2}
\theta(1-\varphi^2)$. Therefore, in the limit $k'\to 0$, the
master equation takes on the following form
\begin{eqnarray}
0&=&-[(\varphi-\varphi^3)P]_\varphi+k(0)P_{\varphi\varphi}- \cr \nonumber \\
&&(k'/2)^{1/2}\rho[A+2\sqrt{2}\varphi^2\delta(\varphi^2-1)]\;,
\end{eqnarray}
where we set $P_t=0$ in the stationary state. The PDF $P$ is
continuous at $\varphi=\pm 1$, but its derivative is not. By
integrating Eq. (24) in the interval $[1-\epsilon,1+\epsilon]$ (or
$[-1-\epsilon, -1+\epsilon]$), one finds
\begin{equation}
\Delta P_\varphi|_{\varphi=\pm 1}=\frac{ (k')^{1/2}\rho}{k(0)}\;.
\end{equation}
In the limit, $k'\to 0$ the derivative of the PDF will also be
continuous. Considering the factor $(k')^{1/2}$ in Eq. (24), we
conclude that, in the strong coupling limit, the unclosed term is
identically zero. This means that there is no anomaly or finite
term in the strong coupling limit in the master equation for the
PDF of the order parameter $\varphi$. The stationary solution of
Eq. (24), in the limit, $k'\to 0$, takes on the following
expression,
\begin{equation}
P_{st}=N\exp\left[\frac{-\varphi^4+2\varphi^2}{4k(0)}\right]\;,
\end{equation}
where the normalization constant is given by
\begin{equation}
\frac{1}{N}=\frac{1}{\sqrt{2}}\exp\left[\frac{1}{8k(0)}\right]
K_{\frac{1}{4}} \left(\frac{1}{8k(0)}\right)\;.
\end{equation}
where the $K_ a (b)$ is the modified Bessel functions. To derive
the moments of $\langle\varphi^n\rangle$ in the stationary state,
we multiply Eq. (25) by $\varphi^n$ and integrate the result over
$\varphi$ to obtain
\begin{eqnarray}
n\langle\varphi^n\rangle-n\langle\varphi^{n+2}\rangle\nonumber\\
+n(n-1)k(0)n\langle\varphi^{n-2}\rangle=0\;.
\end{eqnarray}
Equation (28) is a recursive equation for computing all the
moments in terms of the second-order one. Direct calculation then
shows that,
\begin{equation}
\langle\varphi^2\rangle=-\frac{K_{\frac{1}{4}}
\left(\frac{1}{8k(0)}\right) - K_{\frac{3}{4}}
\left(\frac{1}{8k(0)}\right)}{2 K_{\frac{1}{4}}
\left(\frac{1}{8k(0)}\right)}\;,
\end{equation}
and all the odd moments vanish, $\langle\varphi^{2k+1}\rangle=0$.
Therefore, using Eqs. (28) and (29), we are able to derive all the
moments of the order parameter in the $d-$dimensional Ginzburg -
Landau theory in the strong coupling limit.

\begin{figure}[t]
\epsfxsize=9truecm\epsfbox{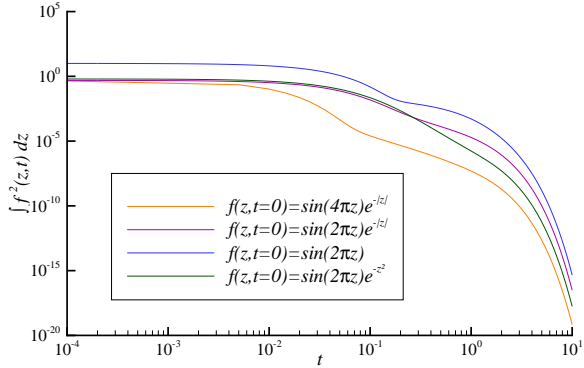} \caption{Time variations of
$\int dz f^2(z,t)$ vs. $t$ with different types of initial
conditions. The results show that $\int dz f^2(z,t)\to 0$ at long
times.}\label{figPDF}
\end{figure}

\section{Master equation for the increments and their PDF tail}

In this section we derive the PDF and the scaling properties of
the moments of the increments, $P(\varphi(x_2)-\varphi(x_1))$, for
the $\varphi^4$-theory in the strong-coupling limit. Defining the
two-point generating function by,
$Z(\lambda)=\langle\Theta\rangle$, where $\Theta$ is defined as
\begin{equation}
\Theta= e^{-i\lambda_1\varphi(x_1,t)-i\lambda_2\varphi(x_2,t)}
\end{equation}
the time evolution of $Z$ is related to that of $\varphi$ by
\begin{eqnarray}
&&Z_t=-i\lambda_1
\langle\varphi_t(x_1,t)e^{-i\lambda_1\varphi(x_1,t)-i\lambda_2\varphi(x_2,t)}
\rangle\nonumber\\
&&-i\lambda_2
\langle\varphi_t(x_2,t)e^{-i\lambda_2\varphi(x_1,t)-i\lambda_2\varphi(x_2,t)}
\rangle\;.
\end{eqnarray}
Substituting $\varphi_t (x_1,t)$ and $\varphi_t (x_2,t)$ from Eq.
(6), the governing equation for the generating function satisfies,
\begin{eqnarray}
&&Z_t=\lambda_1 Z_{\lambda_1}+\lambda_1
Z_{\lambda_1\lambda_1\lambda_1}+\lambda_2 z_{\lambda_2}+\lambda_2
Z_{\lambda_2\lambda_2\lambda_2}\nonumber\\
&&-(\lambda_1^2+\lambda_2^2)k(0)Z-2\lambda_1\lambda_2k(x)Z\;,
\end{eqnarray}
Where, $x=x_1-x_2$, and we have used the fact that in the
strong-coupling limit, the Laplacian term makes no contribution to
the PDF equation (see Appendix II for more details). Moreover, we
have invoked the generalized Novikov's theorem for the two-point
generating function, according to which \cite{16} (see also
Appendix I),
\begin{eqnarray}
&&-i\lambda_1\langle\eta(x_1)
e^{-i\lambda_1\varphi(x_1,t)-i\lambda_2\varphi(x_2,t)}\rangle\nonumber\\
&&-i\lambda_2\langle\eta(x_2)
e^{-i\lambda_1\varphi(x_1,t)-i\lambda_2\varphi(x_2,t)}\rangle\nonumber\\
&&=-(\lambda_1^2+\lambda_2^2)k(0)Z-2\lambda_1\lambda_2k(x)Z\;.
\end{eqnarray}
Fourier transforming Eq. (32), the governing equation of the joint
PDF will be given by
\begin{eqnarray}
&&P_t(\varphi_1,\varphi_2)=
-[(\varphi_1-\varphi_1^3)P]_{\varphi_1}\nonumber\\
&&-[(\varphi_2-\varphi_2^3)P]_{\varphi_2}
+k(0)(P_{\varphi_1,\varphi_1}+P_{\varphi_2,\varphi_2})\nonumber\\
&&+2k(x)P_{\varphi_1,\varphi_2}\;.
\end{eqnarray}
It is useful to change the variables as, $ \varphi_1=(w-u)/2$,
and, $\varphi_2= (w+u)/2$, and, therefore,
$d/d\varphi_1=d/dw-d/du$, and, $d/d\varphi_2= d/dw +d/du$. Now,
Eq. (34) can be written as
\begin{eqnarray}
&& P_t(w,u)=-(wP)_w-(uP)_u +\nonumber\\
&& [\frac{1}{4}(w^3+3wu^2)P]_w+[\frac{1}{4}(u^3+3uw^2)P]_u\nonumber\\
&& 2k(0)(P_{ww}+P_{uu})+2k(x)(P_{ww}-P_{uu})\;.
\end{eqnarray}

To derive the governing equation for the PDF of the increments,
$u=\varphi_2-\varphi_1$, we integrate over $w$ to find that,
\begin{eqnarray}
&& P_t(u)=-(uP)_u+\frac{1}{4}(u^3P)_u\nonumber\\
&& \frac{3}{4}(u\langle w^2\mid u\rangle
P)_u+2[k(0)-k(x)]P_{uu}\;,
\end{eqnarray}
where we have used the fact that the joint PDF $P(w,u)$ can be
written as, $P(w\mid u)P(u)$. It is evident that we cannot derive
a closed equation for the PDF of $u$. Indeed, to determine $P(u)$
we need to know the conditional averaging $\langle w^2\mid
u\rangle$. However, one can derive the tail (both the left and
right ones) of the PDF in the limit, $u\to\infty$. To determine
the tail we note that only near the singularities, in small
separation in space, one finds a large difference in the field
$\varphi$ and, hence, large $u$. On the other hand, near such
points or lines, the field $w$ will be very small. Therefore, in
the limit $u\to\infty$, we can ignore the conditional averaging to
find that,
\begin{equation}
\lim_{u\to\infty}\langle w^2\mid u\rangle\simeq 0\;.
\end{equation}
Therefore, in the limit, $u\to\infty$, we obtain the following
behavior for the tails of the $P(u)$ in the stationary state,
\begin{equation}
P_{st}(\phi_2-\phi_1\to\infty)\sim\exp\left\{-\frac{(\phi_2-\phi_1)^4}
{32[k(0)-k(x)]}\right\}\;.
\end{equation}
To derive the scaling behavior of the moments $\langle u^n\rangle$
one needs to know the entire range of the behavior of the
increments' PDF. Here, we are able to only derive the equation for
the shape of the PDF tails. In the next section, we investigate by
numerical simulation the scaling behavior of the moments $\langle
u^n\rangle$ vs. the separation $x$.

\section{scaling exponents of the moments: numerical simulation}

To calculate numerically the scaling behavior of the moments with
$x$, when $x << 1$, we shall use here the initial-value problem
for the two-dimensional Langevin equation, Eq. (6), in the limit,
$k'\to 0$, when the force is concentrated at discrete times
[17-20]:
\begin{equation}
f(x,y,t)=\sum_j f_j(x,y)\;\delta(t-t_j), \label{kickforce}
\end{equation}
where both the ``impulses'' $f_j(x,y)$ and the ``kicking times''
$t_j$ are prescribed (deterministic or random). The kicking times
are ordered and form a finite or an infinite sequence. The
impulses are always taken to be smooth and acting only at large
scales. The precise meaning that we ascribe to the dynamical
Langevin equation with such a forcing is that, at time $t_j$, the
solution $\varphi(x,y,t)$ changes discontinuously by the amount
$f_j(x,y)$,
\begin{equation}
\varphi(x,y,t_{j+})=\varphi(x,y,t_{j-})+f_j(x,y)\;, \label{discon}
\end{equation}
whereas between $t_{j+}$ and $t_{(j+1)-}$ the solution evolves
according to the unforced $\varphi^4$ equation,
\begin{equation}
\partial_t\varphi= \varphi - \varphi^3 + k'\nabla^2 \varphi.
\label{equnf}
\end{equation}

Without loss of generality, we may assume that the earliest
kicking time is, $t_{j_0}=t_0$, provided that we set,
$f_{j_0}=f_0$, and, $\varphi(x,y,t_{j_0-})=\varphi(x,y,t_{0-})$
for $t<t_0$. Therefore, starting from $t_0$, according to Eq.
(\ref{discon}) we obtain
\begin{equation}
\varphi(x,,y,t_{0+})= \varphi(x,y,t_{0-})+f_0(x,y)\;,
\label{eqdiscon}
\end{equation}
and beyond that up to $t_{1-}$, according to Eq. (\ref{equnf}),
\begin{eqnarray}
&&
\varphi(x,y,t_{1-})=(1+h)\varphi(x,y,t_{0+})-\varphi^3(x,y,t_{0+}),
\nonumber\\
&& h=t_1-t_0 \label{eqalgorithm}
\end{eqnarray}
where, $h=t_1-t_0$.

\begin{figure}[t]
\epsfxsize=9truecm\epsfbox{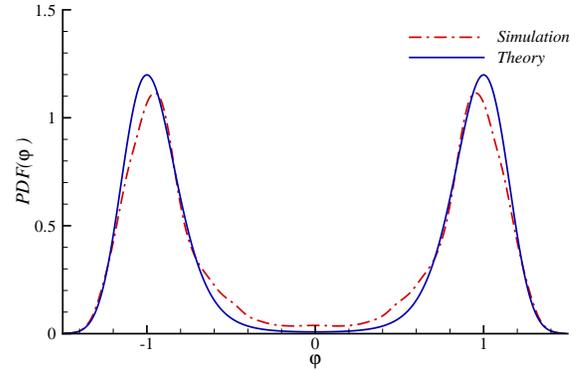} \caption{Exact solution of the
PDF of $\varphi^4$-theory and its comparison with numerical
calculation.}\label{figPDF}
\end{figure}

It is clear that any force $f(x,y,t)$ which is continuously acting
in time can be approximated in this way by selecting the kicking
times sufficiently close. Hereafter, we shall consider exclusively
the case where the kicking is periodic in both space and time.
Specifically, we assume that the force in the $\varphi^4$ equation
is given by
\begin{eqnarray}
f(x,y,t) &=& g(x,y)\sum_{j=-\infty}^{+\infty}
\delta(t-jT),\label{forceperiodic}\\
 g(x,y)&\equiv& -\nabla G(x,y),
\label{defg}
\end{eqnarray}
where $G(x,y)$, the kicking potential, is a deterministic function
of ($x,y$) which is periodic and sufficiently smooth (e.g.,
analytic), and $T$ is the kicking period.

The numerical experiments reported hereafter were made with the
kicking potential $G(x,y)=G_1(x) G_1(y)$, where $G_1(q)$ is given
by
\begin{equation}
G_1(q)= \frac{1}{3}\sin 3q +\cos q, \label{defG}
\end{equation}
and the kicking period, $T=10^{-6}$. The number of collocation
points chosen for our simulations is generally $N_x=10^3$. In Fig.
4 we plot the PDF of $\varphi$ according to Eq. (26), and compare
it with the numerical results. In Fig. 5, the moments of $\varphi$
increments, $\langle|\varphi(x_2+\delta
x)-\varphi(x_1)|^b\rangle$, are calculated numerically as a
function of $x=|x_2-x_1|$ for several values of $b$ (with $0< b <
1$, and $1\leq b $) and its scaling exponents $\xi_b$ for $ x <<
1$ are checked. The results indicate that with good precision
$\langle |\delta\varphi|^b\rangle$ scales with $x$ with an
exponent $1$ for $b> 1$; otherwise, it scales with $x$ with
exponents $\xi_b=b$. Values of $\xi_b$ are given in Fig. 6.

The bi-fractal behavior of the exponents is a consequence of the
presence of the domain walls. Indeed, the structure function,
\begin{equation}
C_b =\langle|\varphi(x_2)-\varphi(x_1)|^b\rangle\;,
\end{equation}
for $b > 0$ behaves, for small $\Delta x=|x_2-x_1|$ as,
\begin{equation}
C_b \sim A_b |\Delta x|^b + A'_b |\Delta x|
\end{equation}
where the first term is due to the regular (smooth) parts of the
order parameter $\varphi$, while the second one is contributed by
the $O(|\Delta x|)$ probability to have a domain wall somewhere in
an interval of length $|\Delta x|$. For $0 < b <1$ the first term
dominates as $|\Delta x|\to 0$, while, for $b>1$ it is the second
term that does so.

\section{Summary}

We studied the domain wall-type solutions in the
$\varphi^4$-theory in the strong-coupling limit, $k'\to 0$, in
which the equation develops singularities. The scaling behavior of
the moments of differences of $\varphi$,
$\delta\varphi=\varphi(x_2)-\varphi(x_1)$, and the PDF of
$\varphi$, i.e., $P(\varphi)$, were all determined. It was shown
that in the stationary state, where the singularities are fully
developed, the relaxation term in the strong-coupling limit leads
to an unclosed term in the equation for the PDF. However, we
showed that the unclosed term can be omitted in the
strong-coupling limit. We proved that to leading order, when
$|x_2-x_1|$ is small, fluctuation of the $\varphi$ field is
intermittent for $b\geq 1$. The intermittency implies that,
$C_b=\langle|\varphi(x_1)-\varphi(x_2)|^b>$ scales as
$|x_1-x_2|^{\xi_b}$, where $\xi_b$ is a constant. It was shown,
numerically, that for the space scale $|x_2 - x_1|$ and $b\geq 1$,
the exponents $\xi_b$ are equal to $1$.

\section{APPENDIX I}

In this appendix we provide a proof of Novikov's theorem. Consider
the general stochastic differential equation with the following
form,
\begin{equation}\label{eq1}
\frac{\partial}{\partial
t}\varphi=-\frac{1}{2}L[\varphi(x,t)]+\eta(x,t)\;.
\end{equation}
where $L$ is an operator acting on $\varphi$, and $\eta$ is a
Gaussian noise with the correlation,
\begin{equation}
\langle\eta(x,t)\eta(x',t')\rangle=k(x-x')\delta(t-t')\;.
\end{equation}
The PDF of the random noise has the following form,
\begin{eqnarray}
&&[d\rho(\eta)]=[d\eta]\times\nonumber\\
&&exp[-\frac{1}{2}\int d^dx d^dx'dt dt'
\eta(x,t)B(x-x')\delta(t-t')\eta(x',t')]\;,\nonumber
\end{eqnarray}
where $B(x-x')$ is the inverse of $k(x-x')$, so that,
\begin{equation}\label{eqortogonal}
\int k(x-x')B(x'-x'')d^d x=\delta(x-x'')\;.
\end{equation}

we write the average of $\eta(x,t)F(\eta)$ over the noise
realization as:
\begin{equation}
\langle\eta(x,t)F(\eta)\rangle=\int
\eta(x,t)F(\eta)[d\rho(\eta)]\;.
\end{equation}
By integrating by parts and using Eq. (\ref{eqortogonal}), one
finds that,
\begin{eqnarray}
&& \langle\eta(x,t)F(\eta)\rangle=\nonumber\\
&&\int d^dx'' dt''
\langle\eta(x,t)\eta(x'',t'')\rangle\langle\frac{\partial
F}{\partial\eta(x'',t'')}\rangle\;.
\end{eqnarray}
Now, let us assume the function $F$ to have the following form,
\begin{equation}
F[\eta]=exp(-i\lambda\varphi(x',t))\;.
\end{equation}
so that one finds,
\begin{equation}
\frac{\partial F}{\partial\eta(x'',t'')}=-i\lambda\frac{\partial
\varphi(x',t)}{\partial\eta(x'',t'')}F[\eta]\;,
\end{equation}
Integrating Eq. (\ref{eq1}) with respect to $t$, we find that,
\begin{eqnarray}
&&\varphi(x',t)=\nonumber\\
&&\varphi(x',t_0)-\frac{1}{2}\int_{t_0}^{t} dt''
L[\varphi(x',t'')]+\int_{t_0}^{t} dt''\eta(x',t'')\;.
\end{eqnarray}
This allows us to show that,
\begin{eqnarray}\label{eq2}
&&\frac{\partial\varphi(x',t)}{\partial\eta(x'',t'')}=\nonumber\\
&&-\frac{1}{2}\int_{t_0}^{t} dt'''\frac{\partial
L[\varphi(x,t''')]}{\partial\eta(x'',t'')}+\delta(x'-x'')\theta(t-t'')\;.
\end{eqnarray}
where, in the limit $t''\rightarrow t$, the first term of the
right-hand side of Eq. (\ref{eq2}) will vanish, and we can write
\begin{eqnarray}
&&\langle\eta(x,t)\exp[-i\lambda\varphi(x',t)]\rangle\nonumber\\
&&=(-i\lambda)k(x-x')\langle\exp(-i\lambda\varphi(x',t))\rangle\;.
\end{eqnarray}
where we used, $\theta(0)=1$

\section{appendix II}

In this Appendix we prove that, for example in Eq. (24), the
relaxation term $k'\nabla^2\varphi$ makes no contribution or
anomaly to the PDF of the increments, in the limit $k'\to 0$.

\begin{figure}[t]
\epsfxsize=9truecm\epsfbox{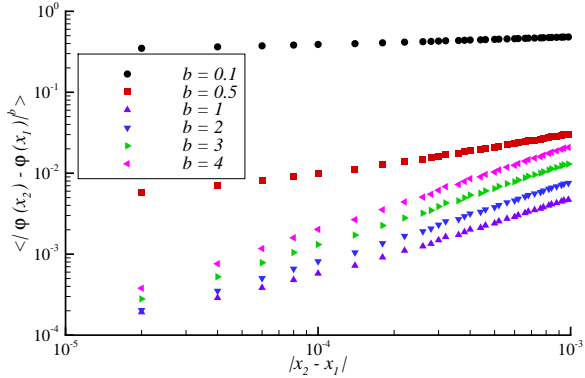} \caption{The moments,
$\langle|\varphi(x_2)-\varphi(x_1)|^b\rangle$, as a function of
$|x_2-x_1|$, obtained via numerical simulation.} \label{figMoment}
\end{figure}
\begin{figure}
\epsfxsize=9truecm\epsfbox{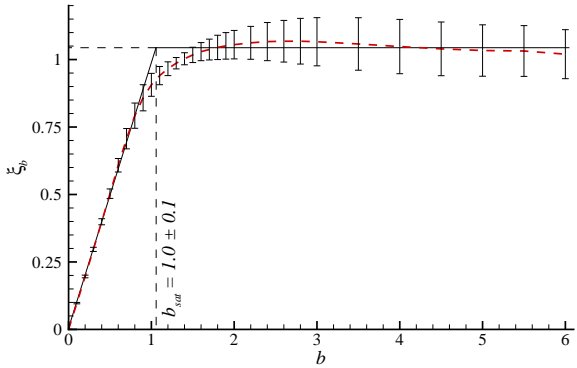} \caption{Scaling exponents
$\xi_b$, checked numerically. The results indicate that, with a
good precision, $\langle|\delta\varphi|^b\rangle$ scales with $x$
with an exponent $1$ for $b>1$ larger than one, and elsewhere
scales with $x$ with an exponent $\xi_b=b$.} \label{figXInt}
\end{figure}

The joint probability distribution $P(\varphi_1,\varphi_2)$
satisfies the following equation,
\begin{eqnarray}
&& P_t(\varphi_1,\varphi_2)=-[(\varphi_1-\varphi_1^3)P]_{\varphi_1}\nonumber\\
&&-[(\varphi_2-\varphi_2^3)P]_{\varphi_2}+k(0)(P_{\varphi_1,\varphi_1}
+P_{\varphi_2,\varphi_2})\nonumber\\
&&+2k(x)P_{\varphi_1,\varphi_2}\nonumber\\
&&-i k'\int \frac{d\lambda_1}{2\pi}\frac{d\lambda_2}{2\pi}
\lambda_1\exp(i\lambda_1\varphi_1+i\lambda_2\varphi_2)\nonumber\\
&&\times\langle\nabla^2\varphi(x_1)
\exp[-i\lambda_1\varphi(x_1,t)-i\lambda_2\varphi(x_2,t)]\rangle\nonumber\\
&&-i k'\int \frac{d\lambda_1}{2\pi}\frac{d\lambda_2}{2\pi}
\lambda_2\exp(i\lambda_1\varphi_1+i\lambda_2\varphi_2)\nonumber\\
&&\times\langle\nabla^2\varphi(x_2)
\exp[-i\lambda_1\varphi(x_1,t)-i\lambda_2\varphi(x_2,t)]\rangle\;.
\end{eqnarray}
The last two terms in Eq. (47) are not closed with respect to the
PDF. Let us then compute the contribution of the unclosed terms.
They can be written as,
\begin{eqnarray}
&&-i k^{'}\int \frac{d\lambda_1}{2\pi}\frac{d\lambda_2}{2\pi}
\lambda_i exp(i\lambda_1\varphi_1+i\lambda_2\varphi_2)\\\nonumber
&&\times<\nabla^2\varphi(x_i)
e^{-i\lambda_1\varphi(x_1,t)-i\lambda_2\varphi(x_2,t)}>\cr
\nonumber\\&& =-i k^{'}(\int
\frac{d\lambda_1}{2\pi}\frac{d\lambda_2}{2\pi}
exp(i\lambda_1\varphi_1+i\lambda_2\varphi_2)\\\nonumber
&&\times<\nabla^2\varphi(x_i)
e^{-i\lambda_1\varphi(x_1,t)-i\lambda_2\varphi(x_2,t)}>)_{\varphi_i}
\nonumber\\&&
=-k'\langle\nabla^2\varphi(x_i)\delta(\varphi_1-\varphi(x_1,t))\\\nonumber
&&\times\delta(\varphi_2-\varphi(x_2,t))\rangle_{\varphi_i}
\end{eqnarray}

Consider one of the terms in the above equation, for example,
$-k'\langle\nabla^2\varphi(x_i)\delta[\varphi_1-\varphi(x_1,t)]
\times\delta[\varphi_2-\varphi(x_2,t)]\rangle_{\varphi_i}$.
Assuming ergodicity, it is written as,
\begin{equation}
= k'\lim_{V\to\infty}\frac{1}{V}\int_V dx_i
dv^i_{d-1}\nabla^2\varphi(x_i)\delta[\varphi_i-\varphi(x_i,t)]
\end{equation}
in the limit, $k'\to 0$ limit, only at the points where we have
singularity this term is not zero. Therefore, we restrict
ourselves to the space near the domain walls,
\begin{eqnarray}
= -k'\lim_{V\to\infty}\frac{1}{V} \sum_{j}\int_{\Omega_{j}}
dx_idv^i_{d-1}(\varphi_{x_ix_i}+\nabla^{2}_{d-1}\varphi)\nonumber\\
\times\delta[\varphi_i-\varphi(x_i,t)]
\end{eqnarray}
where $\Omega_j$ is the space close to the domain walls.
Therefore, Eq. (52), in the limit, $k'\to 0$, is written as,
\begin{equation}
=-k'\lim_{V\to\infty}\frac{1}{V}\sum_j\int_{\Omega_{j}}
dx_idv^i_{d-1}\varphi_{x_ix_i}\delta[\varphi_i-\varphi(x_i,t)]\;.
\end{equation}
Changing the variables from $x_i$ to $z_i$ and integrating over
$dv^i_{d-1}$, one finds,
\begin{eqnarray*}
= -k'\lim_{V\to\infty}
\frac{V_{d-1}}{V}\sum_j\int_{-\infty}^{+\infty} \epsilon
dz_i\frac{1}{\epsilon^2}\varphi_{z_iz_i}
\delta[\varphi_i-\varphi(z_i,t)]\nonumber\\
=-\frac{k'}{\epsilon}\lim_{V\to\infty}\frac{V_{d-1}}{V}\sum_j
\int_{-\infty}^{+\infty}dz_i
\varphi_{z_iz_i}\delta[\varphi_i-\varphi(z_i,t)]\\
\end{eqnarray*}
where $\epsilon=(2k')^{1/2}$. Assuming statistical homogeneity, we
have
\begin{equation}
=-\frac{k'}{\epsilon}\lim_{V\to\infty}\frac{NV_{d-1}}{V}
\int_{-\infty}^{+\infty}dz_i
\varphi_{z_iz_i}\delta[\varphi_i-\varphi(z_i,t)]\\
\end{equation}
where $N$ is number of singular lines. Moreover,
$\frac{k'}{\epsilon}= (k')^{1/2}$, and, $\frac{NV_{d-1}}{V}$ is
the density of the singular lines which, in the limit,
$V\to\infty$, is simply the singularity density $\rho$. Therefore,
\begin{equation}
= -(k')^{1/2}\rho\int_{-\infty}^{+\infty}dz_i \varphi_{z_iz_i}
\delta[\varphi_i-\varphi(z_i,t)]\;.
\end{equation}
In the same way in, for example Eq. (21), by changing the
integration variable from $z_i$ to $\varphi_i$, we calculate the
integral exactly,
\begin{eqnarray*}
=\sqrt{k'/2}\varphi\{A+2\sqrt{2}\phi_i ^2\delta(\phi_i -1)\}
\end{eqnarray*}
where
$A=2\sqrt{2}\theta(1-\phi_i^2)\theta(\phi_i^2-1)-\sqrt{2}\theta(1-\phi_i^2)$.
Therefore, in the limit, $k'\to 0$ the master equation will give
Eq. (35).

\end{document}